%% file: main.tex
\documentclass[10pt,conference]{IEEEtran}
\IEEEoverridecommandlockouts
\usepackage[numbers,sort&compress]{natbib}
\usepackage{amsmath,amssymb,amsfonts}

\usepackage[ruled,vlined,linesnumbered]{algorithm2e}

\usepackage{graphicx}
\usepackage{booktabs}
\usepackage{array}
\usepackage{tabularx}
\usepackage{stmaryrd}
\usepackage{textcomp}
\usepackage{xcolor}
\usepackage[most]{tcolorbox}
\usepackage{amsmath}
\usepackage{amssymb}
\usepackage{booktabs}
\usepackage{listings}
\usepackage{subcaption}
\usepackage{tikz}
\usetikzlibrary{arrows.meta, positioning}
\usepackage{fontspec}
\DeclareFontShape{TU}{TeXGyreTermes}{m}{sc}{<->ssub*TeXGyreTermes/m/n}{}
\usepackage[colorlinks=true,linkcolor=blue,citecolor=blue,urlcolor=blue]{hyperref}
\def\BibTeX{{\rm B\kern-.05em{\sc i\kern-.025em b}\kern-.08em
    T\kern-.1667em\lower.7ex\hbox{E}\kern-.125emX}}

\newcommand{\tool}{\textsf{C2Btor}}
\newcommand{\toolmath}{\text{\tool}}
\newcommand{\btorkw}[1]{\textsf{#1}}

\newtcolorbox{rqanswerbox}{
  enhanced,
  breakable,
  colback=black!5,
  colframe=black!5,
  boxrule=0pt,
  borderline west={2.5pt}{0pt}{black!45},
  left=5pt,
  right=5pt,
  top=4pt,
  bottom=4pt,
  sharp corners
}

\newcolumntype{Y}{>{\raggedright\arraybackslash}X}

\lstdefinestyle{paper-code}{
  basicstyle=\ttfamily\scriptsize,
  columns=fullflexible,
  keepspaces=true,
  breaklines=true,
  frame=single,
  xleftmargin=0pt,
  xrightmargin=0pt,
  aboveskip=2pt,
  belowskip=2pt,
  moredelim=**[is][\color{red!70!black}\bfseries]{@}{@}
}

\newenvironment{paperexample}[1]{%
  \par\smallskip\noindent\textbf{#1.}\ }{\par\smallskip}

\begin{document}

\title{BTOR2-Based C Program Verification via Hardware Model Checking}

\author{
\IEEEauthorblockN{
Xinyu Zhang,
Runxuan Fang,
Ziqun Bao,
Yechuan Xia,
Jianwen Li\textsuperscript{*},
and Geguang Pu
}
\IEEEauthorblockA{
East China Normal University\\
\{westtide,rxfang,52285902029\}@stu.ecnu.edu.cn\\
\{xiaozi465,lijwen2748\}@gmail.com, ggpu@sei.ecnu.edu.cn
}
}

\maketitle

\input{src/0-abstract}

\begin{IEEEkeywords}
Program Verification, BTOR2, Model Checking, Intermediate
Representation
\end{IEEEkeywords}

%%
%% 所有提交的论文必须符合 IEEE 会议论文集模板，具体格式请参见《IEEE 会议论文集格式指南》
%% （标题使用 24 号字体，正文使用 10 号字体；
%% LaTeX 用户必须使用 `\documentclass[10pt,conference]{IEEEtran}`，
%% 无需包含 `compsoc` 或 `compsocconf` 选项）。
%% 请注意，今年采用的是 IEEE 格式，而去年使用的是 ACM 格式。
%% 所有投稿论文正文（包括所有图表、附录等）不得超过10页。
%% 参考文献可额外增加两页。所有投稿必须为PDF格式。
%% 被接收的论文在最终定稿的正文部分可增加一页。
%% 提交的稿件必须严格遵守上述IEEE会议论文集格式规范。
%% 任何对间距、字体大小或其他不符合规范的更改都可能导致稿件被直接拒收，不再进行进一步审核。
%%
\input{src/1-introduction}

\begin{figure*}[t]
\centering
\includegraphics[width=\textwidth]{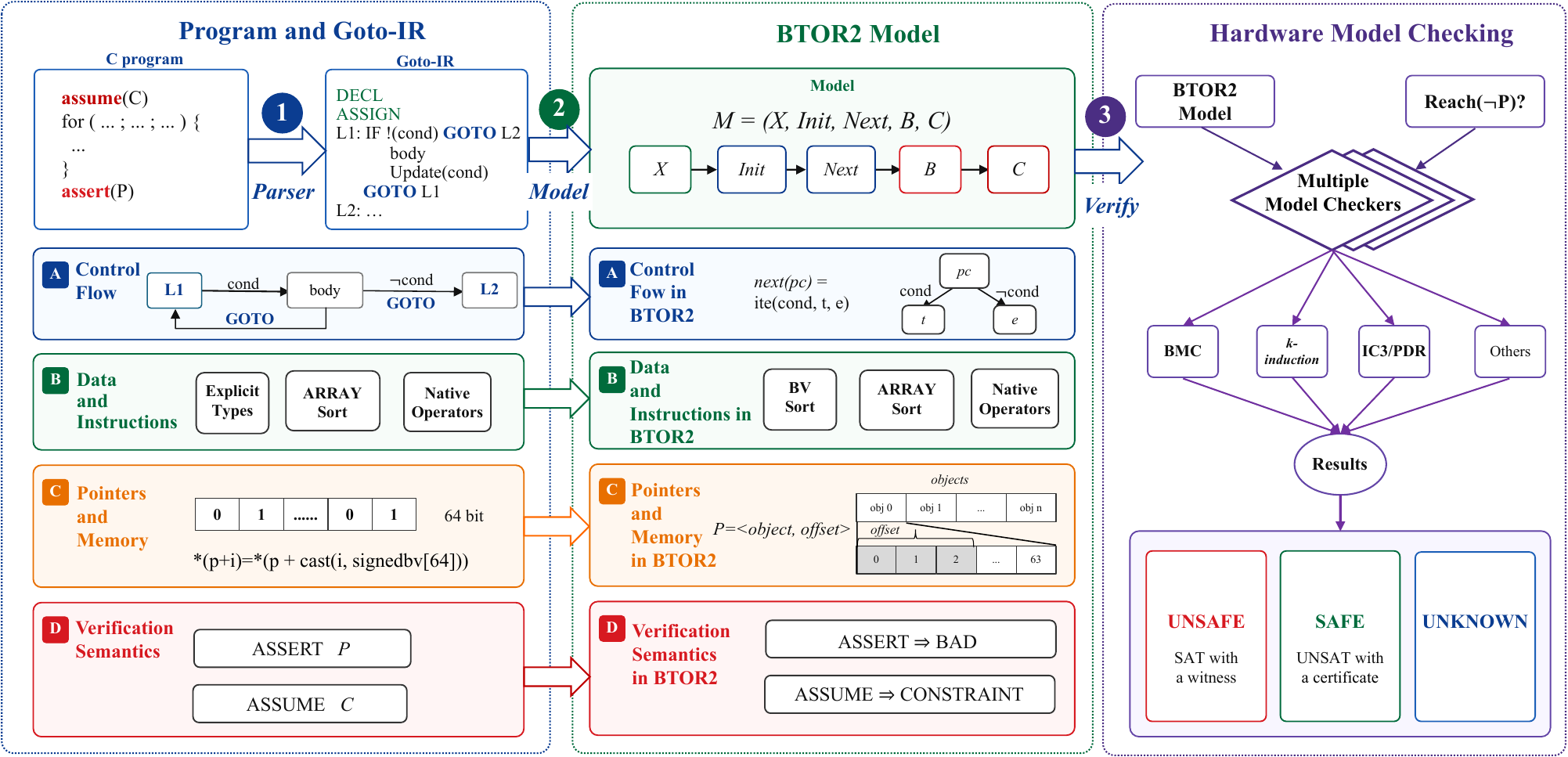}
\caption{Overview of C2Btor.}
\label{fig:c-to-btor2-overview}
\end{figure*}

\input{src/2-related-work}

\input{src/3-background-revised}

\input{src/4-methodology}

\input{src/5-experiment-results}

\input{src/6-conclusion}

\bibliographystyle{IEEEtran}
\bibliography{src/bibliography}
\end{document}

%% file: src/0-abstract.tex
\begin{abstract}
Program verification tools often rely on specific intermediate representations and analysis backends, limiting the reuse of verification algorithms and model checkers across frameworks.
In contrast, hardware model checking has developed a mature backend ecosystem, where standard formats such as BTOR2 support reusable algorithms for counterexample search and inductive safety proving.
Applying these capabilities to C requires translating assertion-based programs into transition systems that hardware model checkers can directly process.

We present \tool{}, a method for encoding such verification tasks into BTOR2 models.
\tool{} uses a program counter to capture control transfers, represents data states and memory objects with bit-vectors and arrays, and maps assumptions and assertion checks into BTOR2 constraints and bad-state properties.

We evaluate \tool{} on SV-COMP C ReachSafety benchmarks and a curated assertion-category benchmark suite, comparing it with representative program verification tools.
\tool{} correctly solves 263 tasks, 101 more than CBMC configured with bounded model checking, and is especially effective on bit-vector benchmarks, where it solves 75.5\% of the tasks with no wrong verdicts.
These results show that the BTOR2 route allows C program verification to benefit from advances in hardware model-checking backends, expanding the available capability for counterexample search, inductive safety proving, and word-level transition-system reasoning.
\end{abstract}

%% file: src/1-introduction.tex
\providecommand{\tool}{\textsf{C2Btor}}
\providecommand{\btorkw}[1]{\textsf{#1}}

\section{Introduction}
\label{sec:introduction}

Hardware model checking has developed a mature ecosystem of automated backends for Reachability-safety analysis.
Through competitions such as the Hardware Model Checking Competition (HWMCC)~\cite{biere2025hardware}, formats such as AIGER~\cite{biere2011aiger} and BTOR2~\cite{niemetz2018btor2} have become stable interfaces, enabling algorithms such as Bounded Model Checking (BMC)~\cite{biere2009bounded}, k-Induction~\cite{jovanovic2016property}, and IC3/PDR~\cite{bradley2011sat,een2011efficient} to be implemented and reused over common inputs.
Although these techniques were developed mainly for hardware transition systems, C program verification has a closely related safety objective: a program is unsafe if some feasible execution reaches an assertion failure, and safe otherwise.
This raises a natural question: Can C reachability-safety tasks benefit from the mature automated backends developed in hardware model checking?

C program verification already has mature intermediate representations and tool ecosystems, as reflected in SV-COMP~\cite{beyer2026evaluating}.
Representations such as LLVM-IR~\cite{merz2012llbmc}, Constrained Horn Clauses (CHCs)~\cite{gurfinkel2022program}, and SMT-LIB~\cite{barrett2010smt} have supported symbolic execution~\cite{king1976symbolic}, abstract interpretation~\cite{afzal2019veriabs}, bounded model checking, and Horn-clause solving~\cite{bjorner2015horn}.
However, these representations are not the task formats directly consumed by HWMCC-style hardware model checkers.
The issue is therefore not the absence of software-verification IRs, but the lack of a semantic encoding layer from assertion-based C verification tasks to hardware model-checking transition systems.

\textbf{This paper addresses this gap by using BTOR2 as the primary target representation.}
The reason is not that BTOR2 replaces lower-level formats such as AIGER, but that it provides a suitable interface for constructing program-level transition systems before optional lowering to bit-level models.
BTOR2 preserves word-level bit-vector and array terms, and provides native constructs for states, transitions, constraints, and bad-state properties.
These features allow the encoding to keep finite-state program data, control-flow locations, assumptions, and assertion failures explicit in the generated model.
AIGER remains useful as a lower-level target for the subset of generated models that can be bit-blasted and discharged by AIGER-based engines.

\textbf{We therefore propose \tool{}, a framework that translates C programs into BTOR2 transition systems and verifies the generated tasks using hardware model checkers.}
To separate C frontend processing from BTOR2-specific semantic encoding, \tool{} follows a three-stage route.
\begin{enumerate}
\item It uses the parser of CBMC~\cite{clarke2004tool} to lower the input C program into Goto-IR\footnote{\url{https://diffblue.github.io/cbmc/group__goto-programs.html}}, a verification-oriented IR with explicit control flow and type information.
\item \tool{} encodes the Goto-IR program into a BTOR2 transition system, including program-counter-based control flow, state and next expressions for data updates, object-level pointer modeling, and the mapping from assumptions and assertions to \btorkw{constraint} and \btorkw{bad} nodes.
\item The generated BTOR2 task is discharged by hardware model-checking backends for counterexample search and safety proving, rather than by CBMC's bounded-model-checking backend.
\end{enumerate}
The main contributions of this paper are as follows:

\begin{itemize}
\item \textbf{A C-to-BTOR2 encoding for a well-defined verification fragment.}
We propose a set of encoding rules that translate assertion-based C verification tasks, via Goto-IR, into BTOR2 transition systems.
The encoding uses an explicit program counter to model control flow, represents data states and memory objects with bit-vectors and arrays, and maps assumptions and assertions uniformly to \btorkw{constraint} and \btorkw{bad} nodes.
This formulation makes explicit the C fragment naturally supported by the BTOR2-based route, including finite-state, bit-vector-friendly programs with structured control flow, nondeterministic inputs, fixed-width integer operations, local arrays, and restricted object-level pointer relations.

\item \textbf{An end-to-end framework for reusing hardware model checking backends.}
We develop \tool{}, a tool framework that connects C reachability-safety tasks, the proposed BTOR2 encoding, and existing hardware model-checking backends.
Given an assertion-based C program, \tool{} generates backend-consumable BTOR2 models and enables different backend algorithms to be applied to the same transition-system representation for counterexample search and safety proving.

\item \textbf{Empirical findings on complementarity and backend algorithm behavior.}
We evaluate \tool{} on SV-COMP C ReachSafety benchmarks and a curated assertion-category dataset, showing that the BTOR2-based route can solve tasks not covered by bounded, SMT-based, or CHC-based software verifiers.
We further analyze the generated model-checking tasks on 390 BTOR2 models and 305 converted AIGER models, showing that backend algorithms play distinct roles in counterexample search and safety proving.
These findings indicate that encoding C verification tasks into BTOR2 allows program verification to benefit from advances in HWMCC-style backend algorithms, thereby expanding the available capability for bug finding and safety proving.
\end{itemize}

The remainder of this paper is organized as follows.
Section~\ref{sec:related-work} discusses related work.
Section~\ref{sec:background} introduces BTOR2, Goto-IR, and the formulation of ReachSafety tasks in hardware model checking.
Section~\ref{sec:methodology} presents the proposed encoding method, including the overall workflow, key translation rules, and semantic boundaries.
Section~\ref{sec:evaluation} reports the experimental design and results.
Section~\ref{sec:conclusion} discusses applicability, limitations, and future work, and concludes the paper.

%% file: src/2-related-work.tex
\section{Related Work}
\label{sec:related-work}

\subsection{C Program Verification}

C program verification has been extensively studied, and SV-COMP has helped
establish a mature ecosystem of tools and benchmarks~\cite{beyer2026evaluating}.
The ReachSafety category spans several paradigms: CBMC~\cite{clarke2004tool}
and ESBMC~\cite{cordeiro2011smt} perform bit-precise bounded model checking;
Ultimate Automizer~\cite{heizmann2018ultimate} uses interpolation and
refinement; CPAchecker~\cite{baier2024software} and
Bubaak~\cite{chalupa2025bubaak} reflect configurable and portfolio
verification; and AISE~\cite{wang2024aise}, Symbiotic~\cite{jonavs2024symbiotic},
and VeriAbs~\cite{afzal2019veriabs} combine abstraction, symbolic execution,
and slicing. CHC-based tools such as Eldarica and TriCera~\cite{hojjat2018eldarica,esen2022tricera}
reduce safety proving to invariant inference and Horn solving~\cite{bjorner2015horn}.

These tools differ in both backend algorithms and intermediate representations.
Goto-IR, LLVM-style representations, SMT encodings, and CHCs expose control
flow, data state, memory, and verification conditions in different ways.
This paper is complementary: instead of building another C verifier over a
software IR, we study whether assertion-based C tasks can be encoded into a
standard hardware model-checking format and solved by hardware backends.

\subsection{Hardware Model Checking, AIGER, and BTOR2}

Hardware model checking is reachability analysis over transition systems with
initial conditions, transition relations, and bad states.
BMC~\cite{biere2009bounded} is effective for bounded counterexample search,
while k-Induction~\cite{jovanovic2016property} and IC3/PDR~\cite{bradley2011sat,een2011efficient}
prove unbounded safety through inductive reasoning.
Recent backends also include CAR-style reachability~\cite{zhang2023accelerate,xia2023searching}
and optimized IC3/PDR variants~\cite{zhou2026ic3}.

Hardware model checking also benefits from common task formats.
AIGER~\cite{biere2011aiger} is a compact bit-level representation used by
model checkers such as SimpleCAR~\cite{li2018simplecar}.
BTOR2~\cite{niemetz2018btor2} is a word-level transition-system format over
bit-vectors, arrays, and state updates; it preserves word-level structure before
bit-blasting~\cite{biere2020tutorial}, and rIC3~\cite{su2025ric3} can process
BTOR2 directly. We use BTOR2 as the primary C-encoding target and AIGER as a
lower-level target for the convertible subset.

\subsection{Bridging Hardware and Software Analysis}

Hardware and software verification start from different artifacts, but many
tasks share the same transition-system core: initial states, state updates, and
error states. Existing work has explored the hardware-to-software direction:
Btor2C~\cite{beyer2023bridging} translates BTOR2 circuits into C programs so
that hardware models can be analyzed by software verifiers.

Our work studies the reverse direction: translating C reachability-safety tasks
into BTOR2 transition systems. Compared with Btor2C, we must recover control
flow, data updates, pointer operations, assumptions, and assertion failures from
C and encode them into a hardware model-checking format.

By targeting BTOR2, our encoding turns C reachability-safety tasks into
word-level transition systems consumed by hardware backends, allowing program
verification to benefit from advances in counterexample search and safety
proving.

%% file: src/3-background-revised.tex
\section{Background}
\label{sec:background}

\subsection{Overview of C2Btor}

Figure~\ref{fig:c-to-btor2-overview} shows the pipeline of C2Btor.
The input is a C program with assertions. \tool{} reuses the CBMC frontend to parse, type-check, and normalize the program into Goto-IR.
We use the following notation throughout the paper:
\[
C \xrightarrow{F_{\mathrm{CBMC}}} G_{\mathrm{GotoIR}}
  \xrightarrow{E_{\toolmath}} M_{\mathrm{BTOR2}}.
\]
Goto-IR lowers structured C control flow into explicit program locations and instructions.
\tool{} then encodes the Goto-IR program into a BTOR2 transition system with a set of encoding rules.
The generated BTOR2 model can then be analyzed by different model-checking backends, such as BMC, k-Induction, and IC3/PDR.
The backend result is interpreted as a program-verification result: SAT means Unsafe and UNSAT means Safe, and an inconclusive run is reported as unknown.

Therefore, 
Section~\ref{sec:bg-program-to-mc} explains the connection between program verification and model checking, and 
Section~\ref{sec:bg-goto-ir} introduces the Goto-IR representation. Section~\ref{sec:bg-btor2} summarizes the BTOR2 concepts used in this work.

\subsection{From Program Verification to Model Checking}
\label{sec:bg-program-to-mc}

In a C program, an assertion expresses a condition that must hold whenever execution reaches the corresponding program point.
SV-COMP ReachSafety tasks use the same idea, often by requiring that an error function such as \texttt{reach\_error} is unreachable~\cite{beyer2026evaluating}.

A transition system provides the target form for this reachability problem.
It contains state variables, initial conditions, and transition updates.
Assertions are represented as bad-state predicates, and assumptions restrict the feasible transitions.
If a bad state is reachable, the program is unsafe and the backend may return a counterexample.
If all bad states are proven unreachable, the program is safe under the generated model; otherwise the result is reported as unknown or timeout.

This is the only model-checking interface required by \tool{}.
The encoder translates Goto-IR control and data updates into BTOR2 \btorkw{state} and \btorkw{next} nodes, and maps verification semantics to \btorkw{bad} and \btorkw{constraint} nodes.
The resulting BTOR2 task can then be analyzed by hardware model-checking backends using BMC, \textit{k}-induction, or IC3/PDR.

\subsection{Goto-IR as a Frontend Representation}
\label{sec:bg-goto-ir}

Generating a transition system directly from C source code is complicated.
C contains structured control flow, implicit conversions, expression evaluation rules, object layout, pointer operations, and library-dependent constructs.
To avoid reimplementing a full C frontend, \tool{} uses the Goto-IR produced by the CBMC frontend as its input representation.
We later write \(G_{\mathrm{GotoIR}}\) simply as \(G\) and abstract it as
\[
G=(L,\Sigma,\mathit{Objects},\mathit{Instructs},\mathit{Verif}).
\]
Here,
\[
L=\{\ell_1,\ell_2,\ldots,\ell_n\}
\]
is the set of Goto-IR locations, where each \(\ell_i\) is the location of one Goto-IR instruction.
\(\Sigma\) is the static symbol table with type information. For example,
\[
\begin{aligned}
\Sigma=\{&
\texttt{int x}:\mathsf{signedbv}[64],\\
&
\texttt{int a[2]}:\mathsf{signedbv}[64][2],\\
&
\texttt{int *p}:\mathsf{signedbv}[64]
\}.
\end{aligned}
\]
\(\mathit{Objects}\) is the collection of program state objects. A valuation of these objects describes the dynamic program state:
\[
\nu\in \mathit{Val}_{\Sigma}(\mathit{Objects}).
\]
Thus, at location \(\ell_i\), a Goto-IR state is
\[
(\ell_i,\nu)\in L\times \mathit{Val}_{\Sigma}(\mathit{Objects}).
\]

\(\mathit{Instructs}\) contains the lowered Goto-IR instructions.
High-level C control flow is represented by assignments and explicit jumps.
\begin{paperexample}{Example 1. Control-Flow Lowering}
A conditional branch instruction \texttt{if (A) \{ B \} else \{ C \}} is
lowered to a conditional goto and two explicit branch blocks:
\begin{lstlisting}[style=paper-code]
L1: if !A goto L_else
L2: B
L3: goto L_end
L_else: C
L_end: ...
\end{lstlisting}
A loop \texttt{for (I; A; U) \{ B \}} is lowered to a condition check and a
back edge:
\begin{lstlisting}[style=paper-code]
L1: I
L2: if !A goto L_end
L3: B
L4: U
L5: goto L2
L_end: ...
\end{lstlisting}
\end{paperexample}

Goto-IR represents pointer operations as explicit typed expressions.
Pointer reads and writes keep the dereference operation visible, while pointer arithmetic and casts expose the bit-vector expression used to compute the accessed address.
\begin{paperexample}{Example 2. Pointer Expressions}
A pointer read such as \texttt{r = *(p + i)} is represented as:
\[
r := *(p+\mathsf{cast}(i,\mathsf{signedbv}[64])).
\]
Similarly a pointer write \texttt{*(p + i) = v} can be:
\[
*(p+\mathsf{cast}(i,\mathsf{signedbv}[64])) := v.
\]
\end{paperexample}

\(\mathit{Verif}\) contains verification-relevant instructions such as \(\mathsf{assert}(p)\) and \(\mathsf{assume}(c)\).
Assertions define bad-state conditions, while assumptions define path constraints.
In summary, Goto-IR gives \tool{} an explicit, typed, and verification-oriented program representation; the rules for translating this representation into BTOR2 are presented in Section~\ref{sec:methodology}.

\subsection{BTOR2 Transition Systems}
\label{sec:bg-btor2}

BTOR2 is a word-level transition-system format for model checking~\cite{niemetz2018btor2}.
It is not a programming language with sequential statements.
Instead, a BTOR2 file is a numbered list of typed definitions.
Each line starts with a unique identifier, and later lines refer to earlier
identifiers to build expressions and transition-system components.
Thus, a BTOR2 model is a graph of sorts, state nodes, input nodes, expression
nodes, and property nodes.
We write the generated BTOR2 transition system as
\[
M_{\mathrm{BTOR2}}=(X,\mathit{Init},\mathit{Next},\mathcal{B},\mathcal{C}).
\]
Here, \(X\) is the set of BTOR2 state nodes, \(\mathit{Init}\) gives their
initial values or constraints, and \(\mathit{Next}\) gives their synchronous
next-state updates.
\(\mathcal{B}\) is the set of bad-state predicates, and \(\mathcal{C}\) is the
set of path constraints.

The following sketch shows the style of BTOR2 lines:
\begin{lstlisting}[style=paper-code]
1  sort bitvec 32
2  sort bitvec 64
29 sort bitvec 1
36 constd 1 4          ; integer constant 4
54 eq 29 pc L_cond     ; pc == loop condition
55 slt 29 i 36         ; i < 4
56 not 29 55           ; !(i < 4)
57 and 29 54 56        ; exit guard
59 and 29 54 55        ; body guard
65 add 2 p offset      ; p + offset
101 ite 1 guard val old; guarded read/update
141 next 1 sum 140     ; next(sum)
200 bad 57             ; reachable error
\end{lstlisting}
This example is schematic: names such as \texttt{pc}, \texttt{i}, and
\texttt{sum} stand for previously defined node identifiers.
The important point is that every computed value is a numbered node, and later
lines build larger expressions by referring to earlier identifiers.

The BTOR2 keywords can be grouped as follows.

\smallskip\noindent\textbf{1) Line and node definitions.}
Each line has a unique id and defines either a \btorkw{sort} or a node.
Node lines include constants such as \btorkw{constd}, persistent variables such
as \btorkw{state}, nondeterministic inputs such as \btorkw{input}, and
expression nodes such as \btorkw{eq}, \btorkw{add}, or \btorkw{ite}.

\smallskip\noindent\textbf{2) Transition symbols.}
\btorkw{state}, \btorkw{init}, and \btorkw{next} define the transition system.
\btorkw{state} declares a variable preserved across steps, \btorkw{init}
constrains its initial value, and \btorkw{next} defines its value in the next
step. In encodings produced by \tool{}, \btorkw{ite} is commonly used inside
\btorkw{next} expressions to encode guarded control-flow and data updates.

\smallskip\noindent\textbf{3) Data symbols.}
Data is typed by \btorkw{sort} lines. The main sorts used here are
\btorkw{bitvec} and \btorkw{array}. Bit-vectors represent integers, Booleans,
program locations, object identifiers, and offsets; arrays represent indexed
program objects or memory-like objects. Constants such as \btorkw{const} and
\btorkw{constd} provide typed literal values.

\smallskip\noindent\textbf{4) Operator symbols.}
BTOR2 provides many typed operators; the complete set is given in the BTOR2
specification~\cite{niemetz2018btor2}. This paper mainly uses Boolean and
bit-vector operators such as \btorkw{not}, \btorkw{and}, \btorkw{eq},
\btorkw{slt}, arithmetic operators such as \btorkw{add}, \btorkw{sub},
\btorkw{mul}, \btorkw{udiv}, conditional \btorkw{ite}, and array
\btorkw{read}/\btorkw{write}. Each operator produces another typed node.

\smallskip\noindent\textbf{5) Verification symbols.}
\btorkw{bad} defines an error-state predicate, and \btorkw{constraint}
restricts feasible traces. They correspond to \(\mathcal{B}\) and
\(\mathcal{C}\) in \(M_{\mathrm{BTOR2}}\). A backend checks whether some
\btorkw{bad} node can become true along a trace satisfying the \btorkw{init},
\btorkw{next}, and \btorkw{constraint} lines.

In summary, BTOR2 provides the target vocabulary for numbered expressions,
states, data, operations, and reachability properties.
Section~\ref{sec:methodology} explains how \tool{} performs this encoding for control flow, data state, pointer-related objects, and verification assertions.

%% file: src/4-methodology.tex
\providecommand{\tool}{\textsf{C2Btor}}

\section{Methodology}
\label{sec:methodology}

\subsection{Overview}
\label{sec:methodology-overview}

The goal of this methodology is not to redefine C semantics. Instead, \tool{}
encodes the control flow, objects, instructions, and verification semantics
already made explicit in Goto-IR into a BTOR2 transition system.
Given the Goto-IR object
$G=(L,\Sigma,\mathit{Objects},\mathit{Instructs},\mathit{Verif})$ defined in
Section~\ref{sec:bg-goto-ir}, we write $O=\mathit{Objects}$,
$I=\mathit{Instructs}$, and $V=\mathit{Verif}$ for compact notation. The
overall mapping is
\[
G=(L,\Sigma,O,I,V)
\to
M=(X,\mathit{Init},\mathit{Next},\mathcal{B},\mathcal{C}).
\]
We organize the encoding rules by the following aspects:
\begin{center}
\footnotesize
\setlength{\tabcolsep}{3pt}
\renewcommand{\arraystretch}{1.12}
\begin{tabular}{@{}>{\raggedright\arraybackslash}p{0.29\columnwidth}
                >{\raggedright\arraybackslash}p{0.35\columnwidth}
                >{\raggedright\arraybackslash}p{0.28\columnwidth}@{}}
\toprule
\textbf{Aspect} & \textbf{Goto-IR source} & \textbf{BTOR2 target} \\
\midrule
Control flow & \(L\), \textsf{GOTO}, \textsf{RETURN} &
\(pc\), \(\mathsf{next}(pc)\) \\
Data state & \(\Sigma,O\) & \(X\), \(\mathit{Init}\) \\
Instruction updates & \(I\) & \(\mathit{Next}\) \\
Pointers and memory & \(O_{\mathit{addr}}\), pointer expressions &
\(X_{\mathit{mem}}\), \textsf{read}, \textsf{write} \\
Verification & \(V\) & \(\mathcal{B},\mathcal{C}\) \\
\bottomrule
\end{tabular}
\end{center}

\smallskip\noindent\textbf{Encoding workflow.}
Algorithm~\ref{alg:c2btor-translation} shows the overall encoding workflow. Its
role is to connect the categories above, not to expose implementation
details.

\begin{algorithm}[t]
\footnotesize
\DontPrintSemicolon
\setlength{\algomargin}{0.65em}
\caption{Symbolic workflow of \tool{}}
\label{alg:c2btor-translation}
\KwIn{\(G=(L,\Sigma,O,I,V)\), where \(L=\{\ell_1,\ldots,\ell_n\}\)}
\KwOut{\(M=(X,\mathit{Init},\mathit{Next},\mathcal B,\mathcal C)\)}
\ForEach{\(\ell_i\in L\)}{
  \(\mathit{node}_i\leftarrow\mathsf{bv}_{64}(i)\);\quad
  \(\mathsf{at}_i\leftarrow(pc=\mathit{node}_i)\) \tcp*[r]{control code and guard}
}
\(X\leftarrow\{pc\}\cup X_{\mathit{data}}(\Sigma,O)\cup X_{\mathit{mem}}(\Sigma,O)\cup X_{\mathit{aux}}\) \tcp*[r]{BTOR2 states}
\(\mathit{Init}\leftarrow\mathit{Init}_{pc}\cup\mathit{Init}_{\mathit{data}}\cup\mathit{Init}_{\mathit{mem}}\) \tcp*[r]{initial states}
\ForEach{\(x\in X\)}{\(U_x\leftarrow\emptyset\) \tcp*[r]{guarded updates}}
\(\mathcal B\leftarrow\emptyset\);\quad \(\mathcal C\leftarrow\emptyset\)\;
\ForEach{\(\ell_i\in L\)}{
  \(\iota_i\leftarrow I[\ell_i]\) \tcp*[r]{instruction at location \(\ell_i\)}
  \((\Delta_i,\delta_i,B_i,C_i)\leftarrow\mathsf{Step}(\iota_i,\mathsf{at}_i,\Sigma,O)\) \tcp*[r]{local encoding}
  \(U_{pc}\leftarrow U_{pc}\cup\{(\mathsf{at}_i,\delta_i)\}\) \tcp*[r]{control-flow update}
  \ForEach{\((x,g,v)\in\Delta_i\)}{
    \(U_x\leftarrow U_x\cup\{(g,v)\}\) \tcp*[r]{data or memory update}
  }
  \(\mathcal B\leftarrow\mathcal B\cup B_i\);\quad
  \(\mathcal C\leftarrow\mathcal C\cup C_i\) \tcp*[r]{properties and constraints}
}
\ForEach{\(x\in X\)}{
  \(\mathit{Next}(x)\leftarrow\mathsf{Fold}(U_x,\mathsf{default}(x))\) \tcp*[r]{one \btorkw{next} per state}
}
\Return \(M=(X,\mathit{Init},\mathit{Next},\mathcal B,\mathcal C)\)\;
\end{algorithm}

\subsection{Control-Flow Encoding}
\label{sec:control-flow-encoding}

The program counter $pc$ is a 64-bit bit-vector state. A BTOR2 line such as
\texttt{2 sort bitvec 64} declares the 64-bit sort; the actual program counter
is a \btorkw{state} node that uses this sort. The value stored in $pc$ is not
a BTOR2 line id. Instead, $pc=i$ means that the current transition executes the
Goto-IR location $\ell_i$.
For compact formulas below, $\mathsf{at}_{\ell_i}$ abbreviates $pc=i$.

The two BTOR2 keywords used most often in this encoding are \btorkw{next} and
\btorkw{ite}. A \btorkw{next} node defines the value of a state in the next
transition step. An expression $\mathsf{ite}(c,a,b)$ means ``if $c$ then $a$
else $b$''. A \btorkw{next} defines how a state changes, while
\btorkw{ite} is only a way to build the right-hand side of a \btorkw{next}
definition.

Let $I_i$ be the Goto-IR instruction at location $\ell_i$. The next value of
$pc$ has four common cases:
\[
pc'=
\begin{cases}
  i+1, & I_i \text{ is sequential},\\
  t_i, & I_i=\mathsf{GOTO}(c_i,\ell_{t_i})\land\mathcal{E}(c_i),\\
  i+1, & I_i=\mathsf{GOTO}(c_i,\ell_{t_i})\land\neg\mathcal{E}(c_i),\\
  pc_{\mathit{halt}}, & I_i\in\{\mathsf{RETURN},\mathsf{END}\}.
\end{cases}
\quad (pc=i)
\label{eq:pc-cases}
\] 
Here, $i+1$ denotes the next executable location in the Goto-IR sequence,
$I_i=\mathsf{GOTO}(c_i,\ell_{t_i})$ denotes a conditional Goto-IR branch at
location $\ell_i$, $c_i$ is its guard, $\ell_{t_i}$ is its target location, and
$t_i$ is the encoded $pc$ value of that target. The expression
$\mathcal{E}(c_i)$ is the BTOR2 encoding of the guard, and $pc_{\mathit{halt}}$
is the terminal location. An unconditional \texttt{GOTO} is treated as the same
rule with a true guard. The local successor for a conditional branch is
\[
  \delta_i=\mathsf{ite}(\mathcal{E}(c_i),t_i,i+1).
\]
BTOR2 requires one \btorkw{next} node for $pc$, so all local control-flow rules
are chained into one expression:
\[
\begin{aligned}
  \mathsf{next}(pc)=
  \mathsf{ite}(pc=0,\delta_0,
  \mathsf{ite}(pc=1,\delta_1,
  \ldots,pc_{\mathit{halt}})).
\end{aligned}
\label{eq:pc-next}
\]
The same pattern is used later for data states: \tool{} collects guarded
candidate updates and folds them into a single one,
\[
\begin{aligned}
  \mathsf{next}(x)=
  \mathsf{ite}(g_1,v_1,\mathsf{ite}(g_2,v_2,\ldots,x)).
\end{aligned}
\label{eq:unified-next}
\]
The final default value preserves the current state when no guard is enabled.

Loops are not unrolled by \tool{}. Back edges in Goto-IR are
encoded directly in $\mathsf{next}(pc)$, so a source-level loop becomes a cycle
in the generated BTOR2 transition system. Whether this cycle is explored using
bounded model checking, $k$-induction, IC3/PDR, or another hardware
model-checking algorithm is a responsibility of the backend. Hence,
\tool{} does not produce a fixed-depth C-to-BMC formula; it produces a
transition system that can be consumed by multiple unbounded or bounded
hardware model checkers.

\subsection{Data State}
\label{sec:data-state}

Data-state rules decide which BTOR2 states are created for program objects and
which initial values are placed in $\mathit{Init}$. For each Goto-IR type
$\tau$, let $W(\tau)$ be its bit-width under the frontend's target-machine
configuration; in practice, this width is usually 32 or 64 bits for machine
integer and pointer-related types.

\smallskip\noindent\textbf{1) Scalars and constants.}
Boolean and integer-like objects, including \texttt{char}, \texttt{short},
\texttt{int}, \texttt{long}, unsigned variants, and \texttt{enum} objects, are
represented as bit-vector states:
\[
  x:\tau \quad\mapsto\quad x:\mathsf{BV}(W(\tau)).
\]
For example, in our encoding, \textbf{bool} is encoded as $\mathsf{BV}(1)$,
while \textbf{int} and \textbf{unsigned int} are encoded as $\mathsf{BV}(64)$.
Signedness does not change the state sort. A signed integer and an unsigned
integer of the same width have the same state type; signedness only selects the
operator used later by an instruction. Constants are generated with the width
required by their expression context.

\smallskip\noindent\textbf{2) Arrays.}
For an array object $a:\tau[N]$, \tool{} uses either scalarization or a BTOR2
array state. A small statically bounded array can be scalarized into element
states:
\[
  a\mapsto\{a_0,a_1,\ldots,a_{N-1}\},
  \qquad
  a_i:\mathsf{BV}(W(\tau)).
\]
For larger arrays, \tool{} uses a native BTOR2 array state:
\[
  a:\tau[N] \quad\mapsto\quad a:\mathsf{Array}(N,\tau).
\]
Here, \(\mathsf{Array}(N,\tau)\) is a paper-level shorthand: \(N\) denotes the
index range of the array and \(\tau\) denotes the element type. In BTOR2, this
state is implemented by a native array sort with a bit-vector index sort and an
element sort of width \(W(\tau)\). BTOR2 \btorkw{read} returns an element, and
\btorkw{write} returns an updated array state. Scalarization avoids array
reasoning for small arrays, whereas native array states avoid expanding large
arrays into many state variables.

\smallskip\noindent\textbf{3) Structures and layout.}
Structures are represented by field projection. For example,
\texttt{struct S \{ int x; unsigned int y; \}; struct S s;} is encoded as one
object-level record whose fields are stored separately:
\[
\begin{aligned}
  s &\mapsto \langle x:s_x,\ y:s_y\rangle,\\
  s.x &\mapsto s_x:\mathsf{BV}(64),\qquad
  s.y \mapsto s_y:\mathsf{BV}(64).
\end{aligned}
\]
The first line records that \(s\) is still one structure object; the second line
shows the BTOR2 state created for each field.
Addressable structures and arrays also receive layout information, such as
object identifiers, field offsets, and element sizes. That layout is used by
the pointer rules, but it does not require every object to be represented as a
byte-addressed memory array.

\smallskip\noindent\textbf{4) Pointer values.}
Pointer-typed variables are also stored as 64-bit bit-vector states. This rule
only says how the pointer value is stored. Pointer arithmetic, dereference
reads, writes, and linked data structures are handled by the object-level
pointer model in Section~\ref{sec:object-level-pointers}.

\smallskip\noindent\textbf{5) Initialization.}
Static zero initialization and explicit initializers are encoded in
$\mathit{Init}$. Verification nondeterminism is not initialization: a call such
as \texttt{\_\_VERIFIER\_nondet\_int()} is an instruction-level input and is
captured in $\mathit{Next}$.

\subsection{Instruction Encoding}
\label{sec:instruction-encoding}

Instruction rules describe how Goto-IR expressions and instructions generate
BTOR2 operators and next-state updates.

\smallskip\noindent\textbf{1) Arithmetic expressions.}
Arithmetic expressions are encoded using native BTOR2 bit-vector operators. For
example, addition, subtraction, multiplication, division, and remainder become
the corresponding word-level operators of the required width. Signedness selects
the signed or unsigned variant where BTOR2 distinguishes them.

\smallskip\noindent\textbf{2) Bitwise, logical, and cast expressions.}
Bitwise operations, logical operations, comparisons, shifts, and conditional
expressions are also encoded with native BTOR2 operators, such as
\btorkw{and}, \btorkw{or}, \btorkw{not}, \btorkw{eq}, \btorkw{slt}, shifts, and
\btorkw{ite}. Casts are encoded as truncation, zero extension, sign extension,
or reinterpretation at the required bit-width.

\smallskip\noindent\textbf{3) Assignments.}
For a simple assignment at location $\ell_i$, written $I_i:x:=e$, \tool{}
encodes the right-hand side as $\mathcal{E}(e)$ and adds one guarded update to
$T_x$:
\[
  T_x\leftarrow T_x\cup\{(pc=i,\mathcal{E}(e))\}.
\]
If this is the only assignment to $x$, the resulting update is
\[
  \mathsf{next}(x)=\mathsf{ite}(pc=i,\mathcal{E}(e),x).
\]
If several locations assign to $x$, all candidate updates are combined by the
folded form in Equation~\ref{eq:unified-next}. Pointer writes and array writes
may add extra guards, such as an index equality or an object-identifier check.

\smallskip\noindent\textbf{4) Lowered branches and loops.}
Conditional control flow has already been lowered by Goto-IR. A source
\texttt{if-else} becomes conditional \texttt{GOTO} instructions, and
\texttt{for}/\texttt{while} loops become the same conditional branches with
back edges. Their effect on $pc$ is encoded by the control-flow rules in
Section~\ref{sec:control-flow-encoding}; they are not separate high-level C
constructs in the BTOR2 model.

\subsection{Pointers and Memory}
\label{sec:object-level-pointers}

\tool{} encodes memory as an object-level memory model instead of one flat byte-addressed array, and each statically resolved addressable object
has an object id, and each pointer is interpreted as an object-offset pair. The
object id selects the logical object; the offset is the byte position inside
that object.

\smallskip\noindent\textbf{1) Addressable objects.}
Memory is stored as a finite object map built from the symbol table:
\[
  \mathcal{M}
  =
  \{\mathsf{id}(o)\mapsto\mathsf{repr}(o)\mid o\in O_{\mathit{addr}}\}.
\]
Here, \(O_{\mathit{addr}}\) contains the addressable objects selected for the
encoding. Their backing representations are the data states introduced earlier:
\[
\begin{array}{rcl}
  \text{scalar }x &\mapsto& \mathsf{BV}\text{ state},\\
  \text{array }a &\mapsto& \text{scalarized elements or a BTOR2 array state},\\
  \text{struct }s &\mapsto& \text{projected field states}.
\end{array}
\]
Only objects that can be statically resolved become backing objects for pointer
accesses.

\smallskip\noindent\textbf{2) Pointer representation.}
A pointer value is modeled as
\[
  p=\langle \mathsf{obj}(p),\mathsf{off}(p)\rangle.
\]
For example,
\[
  \&a[i]=\langle \mathsf{id}(a),i\cdot\mathsf{sizeof}(\texttt{int})\rangle,
  \qquad
  \&s.g=\langle \mathsf{id}(s),\mathsf{fieldoff}(g)\rangle.
\]
The pointer type comes from the Goto-IR expression type and the symbol table.
Layout information comes from the target configuration, including
\(\mathsf{sizeof}\), element size, field offset, alignment, and object bounds.
Thus \(p+1\) preserves the object id and advances the offset by the size of the
pointed-to type:
\[
  p+i =
  \langle
    \mathsf{obj}(p),
    \mathsf{off}(p)+i\cdot\mathsf{sizeof}(*p)
  \rangle.
\]

\smallskip\noindent\textbf{3) Dereference by object dispatch.}
A dereference is resolved by dispatching on the object id and then using the
offset inside the selected backing object:
\[
\mathsf{deref}(p,\tau)=
\begin{cases}
  \mathsf{access}_{a}(\mathsf{off}(p),\tau),
    & \mathsf{obj}(p)=\mathsf{id}(a),\\
  \mathsf{access}_{s}(\mathsf{off}(p),\tau),
    & \mathsf{obj}(p)=\mathsf{id}(s),\\
  \mathsf{unsupported},
    & \text{otherwise}.
\end{cases}
\]
For a read, \(\mathsf{access}\) returns the selected scalar, array element, or
structure field. For a write, the same object guard is used to update the
corresponding backing state.

\smallskip\noindent\textbf{4) Validity and unsupported cases.}
Pointer safety is expressed by object-level predicates:
\[
\begin{aligned}
\mathsf{valid}(p,\tau)\equiv{}&
p\neq 0 \land
\bigvee_{o\in\mathcal{T}(p)}
\bigl(
  \mathsf{obj}(p)=\mathsf{id}(o)\\
&{}\land
  \mathsf{aligned}(\mathsf{off}(p),\tau)\\
&{}\land
  \mathsf{off}(p)+\mathsf{sizeof}(\tau)
  \leq_{\mathsf{u}}\mathsf{sizeof}(o)
\bigr).
\end{aligned}
\]
The predicate checks non-nullness, object match, alignment, and object bounds.
When Goto-IR contains such checks as assertions or equivalent guards, their
violations are translated into BTOR2 bad predicates.
If the target object, offset, or access type cannot be statically represented
by the object-level model, \tool{} rejects the translation instead of emitting
an unsound approximation.

In short, pointers are represented as object-offset pairs rather than concrete
byte addresses. Dereferences are resolved by dispatching on the object id to
finite typed backing objects, while the offset, access type, alignment, and
object bounds determine the precise read or write location.

\subsection{Verification Properties}
\label{sec:assert-assume-encoding}

\tool{} encodes verification as reachability over bad predicates and
path constraints. A Goto assertion \(\texttt{ASSERT}\ c_{\ell}\) at location
\(\ell\) becomes a location-guarded bad predicate:
\[
  b_{\ell}=\mathsf{at}_{\ell}\land\neg \mathcal{E}(c_{\ell}).
\]
Thus an assertion fails only when execution reaches \(\ell\) and the encoded
condition is false. A Goto assumption \(\texttt{ASSUME}\ c_{\ell}\) is not an
error; it is a path filter encoded as a constraint:
\[
  \chi_{\ell}=(\neg\mathsf{at}_{\ell})\lor \mathcal{E}(c_{\ell}).
\]
Multiple assertions and assumptions are collected into \(\mathcal{B}\) and
\(\mathcal{C}\), and explicit runtime checks or error locations are handled by
the same bad-predicate rule when they already appear.

\subsection{Summary}
\label{sec:methodology-summary}

The encoding maintains one consistent chain of symbols across the chapter:
\[
  \ell_i
  \quad\mapsto\quad
  \mathit{node}_i=\mathsf{bv}_{64}(i)
  \quad\mapsto\quad
  \mathsf{at}_i=(pc=\mathit{node}_i).
\]
Control-flow rules populate \(U_{pc}\), instruction rules populate the update
lists \(U_x\) for data and memory states, and verification rules populate
\(\mathcal B\) and \(\mathcal C\). Each update list is folded into one BTOR2
\btorkw{next} expression. The result is the BTOR2 transition system
\(M=(X,\mathit{Init},\mathit{Next},\mathcal B,\mathcal C)\).

%% file: src/5-experiment-results.tex
\providecommand{\tool}{\textsf{C2Btor}}

\section{Evaluation}
\label{sec:evaluation}

\subsection{Experimental Design}
\label{sec:eval-design}

Our experiments focus on four research questions:
\begin{description}[]
  \item[RQ1:] Can \tool{} perform end-to-end verification on supported standard C verification tasks?
  \item[RQ2:] Does the C-to-BTOR2 representation route provide complementary solving capability to existing software-verification routes?
  \item[RQ3:] How do hardware model-checking algorithms perform on the verification tasks generated by \tool{}?
  \item[RQ4:] What properties and C semantics are currently covered by \tool{}, and where are its verification boundaries?
\end{description}

\subsection{Experimental Environment}
\label{sec:eval-setup}

All experiments were conducted on the same server equipped with an Intel Xeon Gold 6132 CPU at 2.60 GHz, 28 cores, and 96 GB memory, running CentOS 7 Core on x86\_64.
Each verification task was limited to 900 s and 16 GB\@.
\tool{} is developed based on CBMC 6.7.1.
The implementation is available in the artifact repository.\footnote{\url{https://anonymous.4open.science}}
We implemented our experimental scripts for execution and resource control, while keeping the result classification consistent with BenchExec-style.\footnote{\url{https://github.com/sosy-lab/benchexec}} Each tool result is normalized as follows:
\begin{itemize}
\item \textbf{Correct verdicts.} TT denotes a \textsc{True} verdict on a true task, and FF denotes a \textsc{False} verdict on a false task. The correct-verdict rate is \((TT+FF)/N\).
\item \textbf{Wrong verdicts.} TF is a false alarm, namely a \textsc{False} verdict on a true task; FT is a missed bug, namely a \textsc{True} verdict on a false task. The wrong-verdict rate is \((TF+FT)/N\).
\item \textbf{Inconclusive outcomes.} \textsc{Unknown}, OOT, and OOM mean that the tool does not return a Boolean verdict within the configured resource limit. The inconclusive rate is \((\textsc{Unknown}+OOT+OOM)/N\).
\item \textbf{Abnormal outcomes.} FAIL denotes an unsupported-input failure before solving, while ERR denotes a crash. These outcomes are reported separately when needed because they usually reflect semantic coverage boundaries or implementation robustness problems rather than ordinary backend verdicts.
\end{itemize}

\begin{figure}[!t]
\centering
\includegraphics[width=\columnwidth]{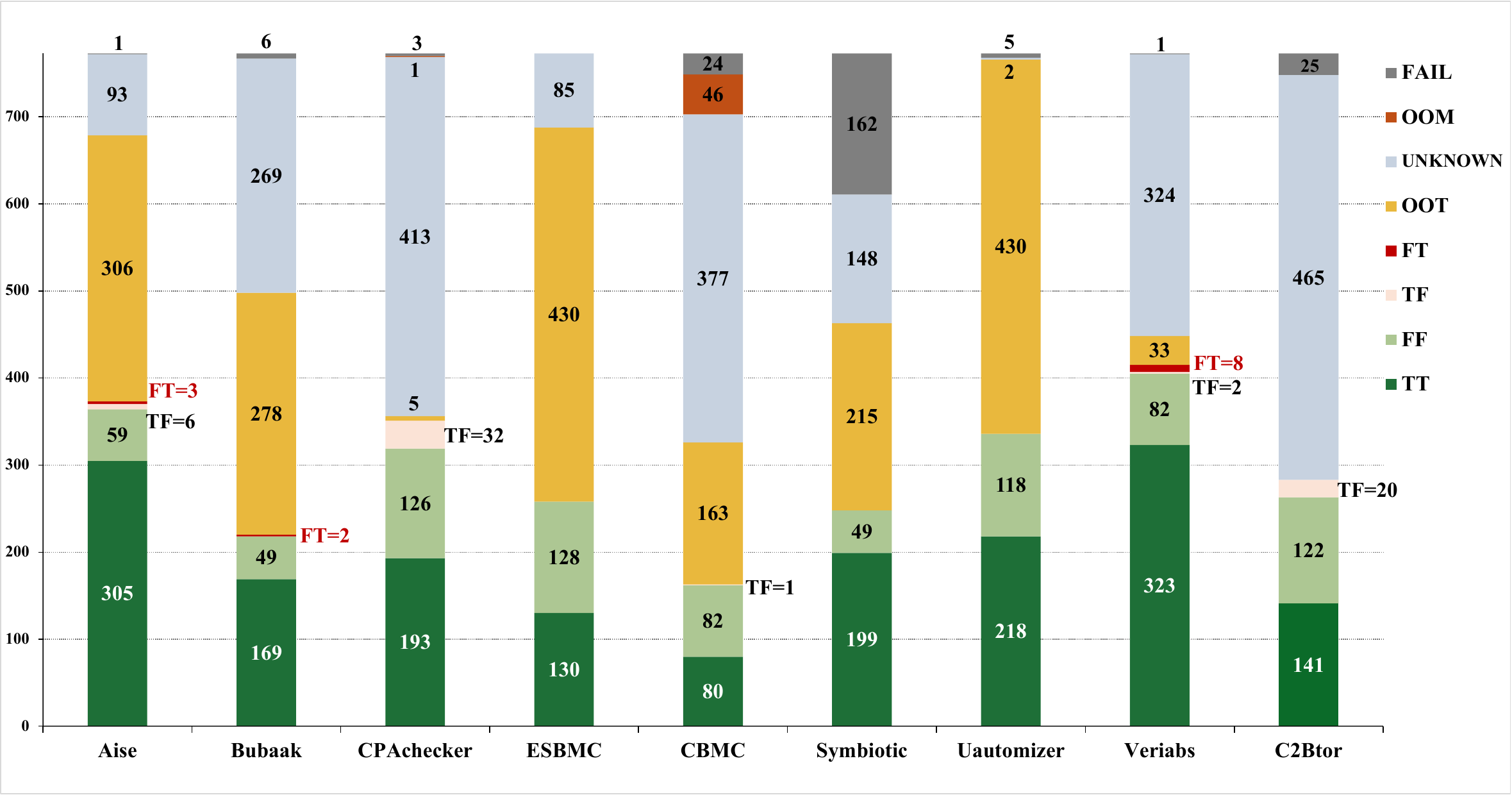}
\caption{Comparison of \tool{} with other program verifiers on SV-COMP benchmarks.}
\label{fig:rq1-result-distribution}
\end{figure}

\begin{figure}[!t]
  \centering
  \includegraphics[width=\columnwidth]{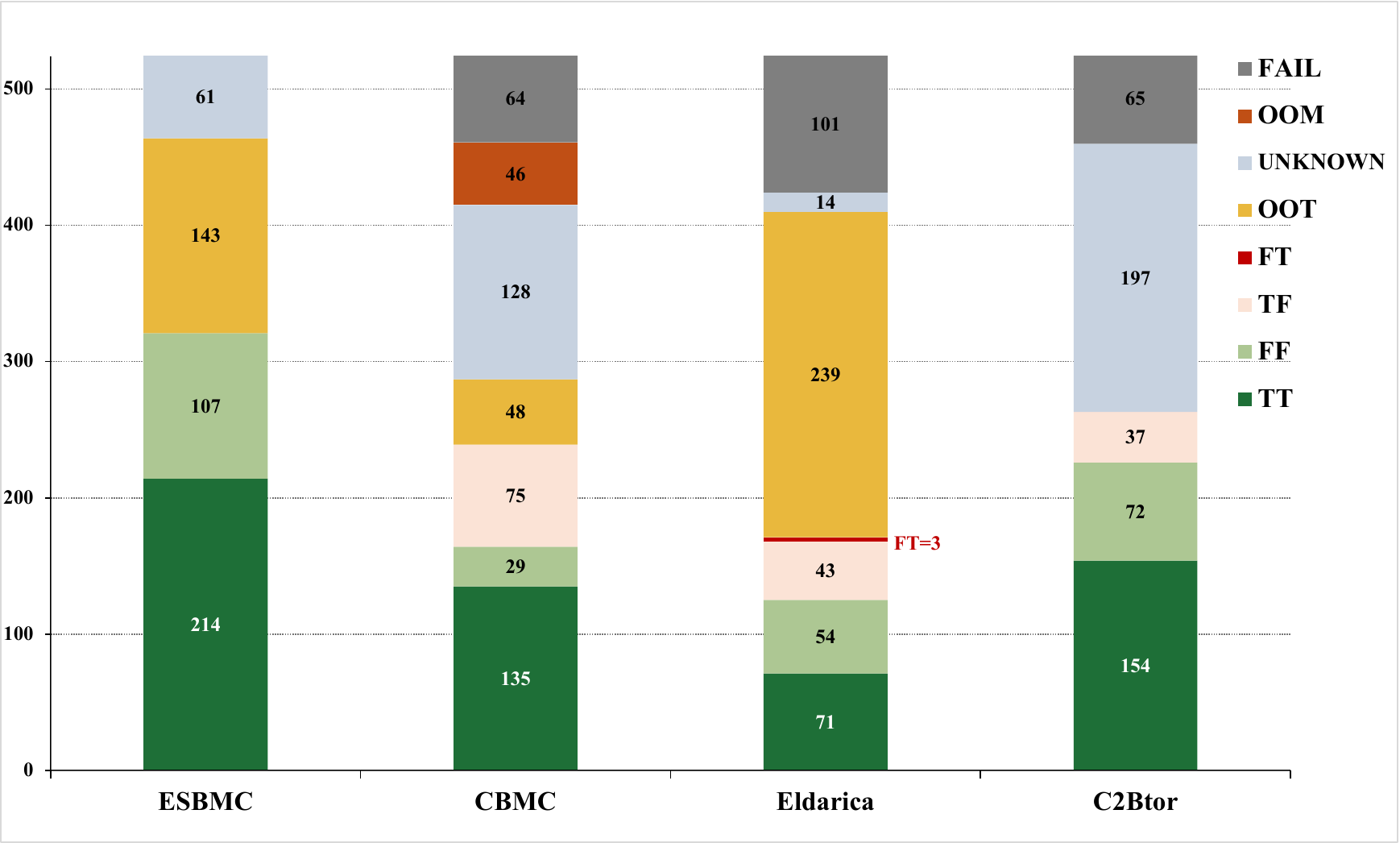}
  \caption{Comparison of assertion verification and expressiveness on a curated benchmark set.}
  \label{fig:rq2-result-distribution}
\end{figure}

\begin{table}[!t]
\centering
\caption{Percentage comparison of \tool{} with other program verifiers on SV-COMP benchmarks (Correct / Wrong / Inconclusive, \%).}
\label{tab:rq1-category-compressed}
\scriptsize
\setlength{\tabcolsep}{2.5pt}
\renewcommand{\arraystretch}{1.08}
\resizebox{\columnwidth}{!}{%
\begin{tabular}{lccccc}
\toprule
\textbf{Tool} & \textbf{Array} & \textbf{BV} & \textbf{Loop} & \textbf{All} & \textbf{TF/FT} \\
 & \textbf{($N=412$)} & \textbf{($N=49$)} & \textbf{($N=312$)} & \textbf{($N=773$)} & \textbf{(all)} \\
\midrule
AISE & 22.3/2.2/75.2 & 67.3/\textbf{0.0}/32.7 & \textbf{76.6}/\textbf{0.0}/23.4 & 47.1/1.2/51.6 & 6/3 \\
Bubaak & 13.1/0.2/85.9 & 38.8/\textbf{0.0}/61.2 & 46.5/0.3/52.2 & 28.2/0.3/70.8 & 0/2 \\
CPAchecker & 17.5/7.8/74.3 & \textbf{89.8}/\textbf{0.0}/\textbf{10.2} & 65.1/\textbf{0.0}/34.6 & 41.3/4.1/54.2 & 32/0 \\
ESBMC & 19.2/\textbf{0.0}/80.8 & 71.4/\textbf{0.0}/28.6 & 46.2/\textbf{0.0}/53.8 & 33.4/\textbf{0.0}/66.6 & 0/0 \\
CBMC-BMC-200 & 9.2/\textbf{0.0}/86.9 & 67.3/\textbf{0.0}/32.7 & 29.2/0.3/67.9 & 21.0/0.1/75.8 & 1/0 \\
Symbiotic & 16.0/\textbf{0.0}/83.5 & 73.5/\textbf{0.0}/24.5 & 46.8/\textbf{0.0}/\textbf{2.2} & 32.1/\textbf{0.0}/47.0 & 0/0 \\
UAutomizer & 19.2/\textbf{0.0}/80.3 & 71.4/\textbf{0.0}/26.5 & 71.2/\textbf{0.0}/28.2 & 43.5/\textbf{0.0}/55.9 & 0/0 \\
VeriAbs & \textbf{45.4}/1.7/\textbf{52.7} & 0.0/\textbf{0.0}/100.0 & 69.9/1.0/29.2 & \textbf{52.4}/1.3/\textbf{46.2} & 2/8 \\
\textbf{C2Btor (Ours)} & \textbf{17.7/4.6/76.2} & \textbf{75.5/0.0/24.5} & \textbf{49.0/0.3/44.6} & \textbf{34.0/2.6/60.2} & \textbf{20/0} \\
\bottomrule
\end{tabular}%
}
\end{table}

\begin{table*}[t]
\centering
\caption{Assertion verification and expressiveness comparison on the curated benchmark set. For each baseline tool, entries report Tool solved / Tool-only / \tool{}-only tasks; the last column reports tasks solved by \tool{}.}
\label{tab:lab2-assert-target-overlap}
\scriptsize
\setlength{\tabcolsep}{3pt}
\renewcommand{\arraystretch}{1.08}
\resizebox{\textwidth}{!}{%
\begin{tabular}{@{}>{\raggedright\arraybackslash}p{0.16\textwidth}
                >{\raggedright\arraybackslash}p{0.30\textwidth}
                r c c c c@{}}
\toprule
\textbf{Assertion target} &
\textbf{Representative tasks} &
\textbf{N} &
\textbf{ESBMC} &
\textbf{CBMC} &
\textbf{Eldarica} &
\textbf{\tool{} (Ours)} \\
\midrule
Functional correctness &
Tree/Array: arithmetic, search, sort, and data-structure operation checks &
324 &
133 / 13 / 2 &
87 / 7 / 42 &
71 / 2 / 53 &
122 \\

Pointer consistency &
Lists/Heap: \texttt{next}/\texttt{prev} pointers, aliasing, and nullness checks &
181 &
168 / 70 / 4 &
70 / 15 / 47 &
40 / 16 / 78 &
102 \\

Boundary safety &
Stack/Queue/Array: capacity and bounds checks &
20 &
20 / 18 / 0 &
8 / 8 / 2 &
15 / 13 / 0 &
2 \\
\midrule
Overall &
Assertion-category dataset &
525 &
321 / 101 / 6 &
165 / 30 / 91 &
126 / 31 / 131 &
226 \\
\bottomrule
\end{tabular}%
}
\end{table*}

\subsection{RQ1: Verification Capability on SV-COMP Benchmarks}
%% Dataset, comparison tools, figures, tables, and metrics.
\textbf{Dataset.}
To evaluate verification capability on standard benchmarks, we use 773 C
ReachSafety tasks from SV-COMP 2026, covering Array, BV, and Loop programs.

\textbf{Comparison tools.}
The comparison covers three groups:
\begin{itemize}
  \item Top tools in the SV-COMP C ReachSafety category\footnote{\url{https://sv-comp.sosy-lab.org/2026/results/results-verified/}}: CPAchecker, ESBMC with \texttt{--kind}, and Symbiotic;
  \item SOTA tools on the Array and Loop subsets: AISE, Bubaak, VeriAbs, and UAutomizer;
  \item Baseline: CBMC with \texttt{--bmc --unwind 200}.
\end{itemize}
The above tools are run with their corresponding SV-COMP tool configurations.\footnote{\url{https://gitlab.com/sosy-lab/benchmarking/fm-tools/-/tree/main/data}}

%% Overall analysis.
Figure~\ref{fig:rq1-result-distribution} shows that \tool{} generates BTOR2 models for 748 out of 773 tasks.
Model generation is reported separately from verification correctness: \tool{} correctly solves 263 tasks, 101 more than the bounded CBMC baseline, while the remaining generated cases may still be inconclusive or wrong-verdict outcomes.
This shows that the BTOR2 route can benefit from backend model-checking algorithms beyond bounded unwinding.
All wrong verdicts of \tool{} are TF false alarms, which can be further inspected through counterexample or witness validation; more importantly, no FT missed bug is observed, whereas three comparison tools report at least one FT case.

%% Strength: BV tasks.
Table~\ref{tab:rq1-category-compressed} reports the same experiment as category-level Correct / Wrong / Inconclusive percentages, with $TF$ for false alarms and $FT$ for missed bugs. The results show that \tool{} is strongest on \textbf{BV tasks}, where it correctly solves 75.5\% of the benchmarks, reports no wrong verdicts, and is second only to CPAchecker.
This advantage is not captured by overall solved rate alone: VeriAbs has the highest overall solved rate in Table~\ref{tab:rq1-category-compressed}, but solves no BV task in this experiment.
The contrast reflects different method strengths.
VeriAbs mainly extends bounded model checking with abstract acceleration and k-induction for large or unknown loop bounds, whereas BV programs match the BTOR2 route more directly: both BTOR2 and hardware model checkers natively support fixed-width bit-vector operations and transition-system structure.
Thus, bit-vector-heavy programs fit the target representation more naturally than memory-intensive programs.

%% Limitation: Array tasks.
\textbf{Array tasks} are the main precision and scalability bottleneck.
\tool{} correctly solves 17.7\% of the Array benchmarks, which is close to CPAchecker but clearly below the strongest tools in this category.
At the same time, Array accounts for 19 of the 20 TF false alarms and has a high inconclusive rate of 76.2\%.
Among the three benchmark categories, Array tasks are the most difficult for the current BTOR2 route: they tend to induce larger bit-level encodings and more OOT outcomes.
This concentration suggests that array-heavy programs stress both the object/array encoding and the backend solver, making this category the current boundary of \tool{}'s accuracy and solving effectiveness.

%% Intermediate behavior: Loop tasks.
\textbf{Loop tasks} show a more balanced result.
\tool{} correctly solves 49.0\% of the Loop benchmarks, outperforming Bubaak and the bounded CBMC baseline.
The low wrong-verdict rate of 0.3\% suggests that the program-counter-based transition relation can handle a meaningful subset of standard loop safety tasks.
Manual inspection of abnormal outcomes indicates that these cases are mainly caused by loop tasks combined with array or memory features, where unsupported frontend cases and backend difficulty overlap.

\begin{rqanswerbox} 
\textbf{Answer to RQ1:}
\tool{} establishes an effective verification route. It generates BTOR2 models
for 96.7\% of the tasks and obtains correct verdicts for 263 tasks, while
other cases remain inconclusive or become TF false alarms. It also performs
well on BV tasks, correctly solving 75.5\% of them without wrong verdicts.
This confirms the effectiveness of the \tool{} modeling route and its
advantage on BV tasks.
\end{rqanswerbox}

%% RQ2: assertion-target complementarity.
\subsection{RQ2: Complementarity across Verification Encodings}

%% Dataset and tool selection.
\textbf{Dataset.}
To evaluate whether the \tool{} route provides solving behavior complementary
to existing verification encodings, especially on what kinds of assertions it
can verify, we combine self-constructed programs with benchmarks from the
TriCera dataset~\cite{esen2022tricera}\footnote{\url{https://zenodo.org/records/5831003}},
yielding a curated assertion benchmark set of 525 tasks.

\textbf{Verification routes.}
There are four representative routes:
\begin{itemize}
  \item ESBMC: LLVM-IR and SMT solving;
  \item CBMC: Goto-IR and SAT solving;
  \item Eldarica: CHC encoding with CHC solving;
  \item \tool{}: BTOR2 encoding with btor-ic3 model checker.
\end{itemize}

From the overall results in Table~\ref{tab:lab2-assert-target-overlap}, \tool{} correctly solves 226 out of 525 tasks, showing that it already provides non-trivial end-to-end solving capability. Compared with ESBMC, the additional gain of \tool{} is limited, suggesting that the LLVM-IR/SMT-based route still provides stronger overall coverage in C semantic parsing, program modeling, and backend verification. However, compared with CBMC and Eldarica, \tool{} solves more tasks correctly. This is because the former are respectively constrained by a single \texttt{--bmc} configuration or a CHC/CEGAR solving route, whereas \tool{} can reuse a broader set of BTOR2 backend algorithms. This indicates that one advantage of the BTOR2 transition-system representation is its ability to host diverse model-checking algorithms.

This complementarity is not uniformly distributed across different assertion targets. For functional-correctness tasks, \tool{} correctly solves 122 tasks, close to the 133 tasks solved by ESBMC\@. This category is the largest one and mainly includes return-value comparisons, local numeric relations, array-element relations, and loop postcondition assertions. Therefore, in this category, \tool{} mainly provides comparable coverage rather than substantial unique solving capability. For the currently supported restricted pointer assertions, \tool{} provides more complementary value relative to CBMC and Eldarica, because pointer relations can often be finitized into object identifiers, field relations, or index relations, and further reduced to bit-vector representations. This matches the bit-precise transition-system representation of BTOR2. In contrast, boundary-safety tasks are not the current strength of \tool{}. These tasks usually involve array bounds, capacity constraints, and range relations inside loops, and therefore more directly expose the coverage boundary of the current translation implementation.

Overall, \tool{} can reuse BTOR2 backend model-checking algorithms, which is one of its advantages over CBMC and Eldarica. For boundary-capacity checks, complex heap/list semantics, and tasks requiring stronger global invariant reasoning, the current method still has room for improvement. At the same time, \tool{} provides an additional solving route for a subset of finite and explicit-state C safety tasks, especially local functional assertions, relational assertions, and assertions expressible as bit-vector reachability.

\begin{rqanswerbox}
\textbf{Answer to RQ2:}
Yes. \tool{} provides a distinct and useful verification route that complements
existing software-verification routes. The BTOR2 route solves tasks that are
not solved by CBMC-BMC and Eldarica, showing independent solving capability.
This complementarity is most visible on functional-correctness and restricted
pointer-consistency assertions, where local value relations, array-element
relations, return-value assertions, and object-level pointer relations can be
effectively represented as finite bit-vector transition systems.
\end{rqanswerbox}

%% BTOR2 and AIGER backend comparison.
\subsection{RQ3: Backend Algorithms on Generated Hardware Models} 
\textbf{Dataset.} 
To evaluate the capability and task sensitivity of backend hardware model-checking algorithms, we start from 390 validated BTOR2 models generated by \tool{} and convert them to AIGER through the \texttt{btor2aiger} tool~\cite{biere2025hardware,niemetz2018btor2}, obtaining a second dataset of 305 AIGER models.
The remaining 85 models are excluded because their array structures are not supported by the tool.

\textbf{Model-checking tools and algorithms.}
We compare the following tools and algorithm configurations:
\begin{itemize}
  \item \texttt{rIC3-BMC}~\cite{su2025ric3}: \texttt{--bmc} for AIGER and \texttt{--wlbmc} for BTOR2;
  \item \texttt{rIC3-k-Induction}: \texttt{--kind} for AIGER and \texttt{--wlkind} for BTOR2;
  \item \texttt{rIC3-IC3}: \texttt{--ic3} for both AIGER and BTOR2;
  \item \texttt{simpleCAR-IC3}~\cite{li2018simplecar,xia2023searching}: \texttt{--ic3} for both AIGER and BTOR2;
  \item \texttt{super\_prove}~\cite{mishchenko2012using}: portfolio-based unbounded model checking on AIGER.
\end{itemize}
For this backend-level experiment, TT denotes a true-label task proved safe by the solver, corresponding to a solver-level \textsc{UNSAT} result.
FF denotes a false-label task for which the solver finds a counterexample, corresponding to a solver-level \textsc{SAT} result.

\begin{table}[!t]
\centering
\caption{Model-checking tool and algorithm comparison on the BTOR2 dataset.}
\label{tab:rq3-BTOR2-backends}
\scriptsize
\setlength{\tabcolsep}{2.5pt}
\renewcommand{\arraystretch}{1.08}
\resizebox{\columnwidth}{!}{%
\begin{tabular}{llrrrrrr}
\toprule
\textbf{Backend} & \textbf{Correct (TT/FF)} & \textbf{TF} & \textbf{FT} & \textbf{OOT} & \textbf{OOM} & \textbf{Unknown} & \textbf{FAIL} \\
\midrule
\texttt{rIC3-BMC} & 69 (0/69) & 0 & 0 & 211 & 0 & 0 & 110 \\
\texttt{rIC3-k-Induction} & 177 (108/69) & 0 & 0 & 103 & 0 & 0 & 110 \\
\texttt{rIC3-IC3} & 253 (177/76) & 0 & 0 & 33 & 0 & 0 & 104 \\
\texttt{simpleCAR-IC3} & 230 (170/60) & 0 & 0 & 23 & 0 & 0 & 137 \\
\bottomrule
\end{tabular}%
}
\end{table}

\begin{table}[!t]
\centering
\caption{Model-checking tool and algorithm comparison on the AIGER dataset.}
\label{tab:rq3-aiger-backends}
\scriptsize
\setlength{\tabcolsep}{2.5pt}
\renewcommand{\arraystretch}{1.08}
\resizebox{\columnwidth}{!}{%
\begin{tabular}{llrrrrrr}
\toprule
\textbf{Backend} & \textbf{Correct (TT/FF)} & \textbf{TF} & \textbf{FT} & \textbf{OOT} & \textbf{OOM} & \textbf{Unknown} & \textbf{FAIL} \\
\midrule
\texttt{rIC3-BMC} & 78 (0/78) & 0 & 0 & 64 & 163 & 0 & 0 \\
\texttt{rIC3-k-Induction} & 212 (134/78) & 0 & 0 & 93 & 0 & 0 & 0 \\
\texttt{rIC3-IC3} & 268 (190/78) & 0 & 0 & 37 & 0 & 0 & 0 \\
\texttt{simpleCAR-IC3} & 275 (198/77) & 0 & 0 & 30 & 0 & 0 & 0 \\
\texttt{super\_prove} & 276 (200/76) & 0 & 0 & 26 & 2 & 0 & 1 \\
\bottomrule
\end{tabular}%
}
\end{table}

Table~\ref{tab:rq3-BTOR2-backends} shows a clear division of labor among BTOR2 backend algorithms between safety proving and counterexample finding.
For tasks that require proving safety, BMC can only search for counterexamples within a bounded unfolding depth and cannot establish unbounded safety.
In contrast, k-induction and IC3/PDR can construct inductive safety evidence, with \texttt{rIC3-IC3} proving the largest number of true-label tasks.
On BTOR2 models generated by \tool{}, word-level BMC is also sensitive to array sorts, memory objects, and control-flow unfolding, which can quickly enlarge the generated formulas and lead to many OOT or FAIL outcomes.
IC3/PDR does not rely only on depth-based unrolling; instead, it advances the reachable state space in a property-directed way, which explains why it covers more safety proofs while still finding counterexamples.

Table~\ref{tab:rq3-aiger-backends} shows that the AIGER subset has a higher solving ratio.
On this bit-level subset, counterexample finding is no longer the main source of difference, since most configurations find a similar number of false-label cases.
The main separation comes from safety proving: IC3-style backends with stronger simplification, induction, and portfolio strategies prove more safe tasks.
This indicates that optimizations in hardware model-checking algorithms can further improve the verification capability of models generated by \tool{}.
However, this result must be interpreted together with the input scope: the AIGER subset excludes array-related models, so its higher solving ratio is not a full replacement for complete BTOR2 coverage.

Overall, the models generated by \tool{} can indeed reuse hardware model-checking backends, but backend effectiveness depends on both the verification objective and the target format.
The advantage of BTOR2 is that it preserves word-level expressiveness and can model array-related programs, although current BTOR2 backends still face input-compatibility and tool-robustness bottlenecks.
After conversion to AIGER, \tool{} can benefit from more mature bit-level backends, but this comes at the cost of losing array-structure coverage.

\begin{rqanswerbox}
\textbf{Answer to RQ3:}
Hardware model-checking algorithms are effective on the verification tasks
generated by \tool{}, and their effects are algorithm-specific. BMC is useful
for exposing counterexamples, while induction- and IC3/PDR-style algorithms
are more effective at proving program safety. The results further show that
progress in hardware model checking can directly improve the verification
capability of C programs translated by \tool{}.
\end{rqanswerbox}

\subsection{RQ4: Semantic Coverage and Verification Boundaries}

\begin{rqanswerbox}
\textbf{Answer to RQ4:}
\tool{} has a clear and useful effective scope: finite-state, bit-precise,
assertion-based reachability safety.
This scope already supports meaningful C verification tasks with explicit
control flow, local data reasoning, and restricted object-level pointer
relations, while also identifying concrete targets for extending BTOR2-based C
verification.
\end{rqanswerbox}

\tool{} encodes an assertion violation as a BTOR2 \btorkw{bad} state, and
verification checks whether this state is reachable in the generated transition
system.
This formulation naturally matches C programs whose behavior can be represented
by explicit control-flow transitions and fixed-width data updates.
The current encoding supports control flow with branches and loops,
nondeterministic inputs, fixed-width integer operations, assumptions and
assertions, local array accesses, and restricted object-level pointer
relations.
The preceding results show that this semantic fragment is practically useful:
\tool{} provides end-to-end verification capability on standard ReachSafety
tasks, behaves particularly well on BV-heavy programs and local functional
assertions, and generates BTOR2/AIGER models that can reuse hardware
model-checking algorithms.

The experiments also make the current boundary concrete.
Array-intensive programs are the main stress cases, because array updates,
symbolic indices, and loops can produce large word-level transition relations
and shift more memory reasoning to the backend.
Programs involving complex aliasing, dynamic heap manipulation, complex
standard-library behavior, floating-point arithmetic, or full C
undefined-behavior semantics require stronger frontend modeling than the
current prototype provides.
These cases point to the next extension targets: more compact memory encodings,
richer heap and alias modeling, library summaries, and broader C semantic
coverage.

Overall, \tool{} already provides a practical hardware-model-checking route for
a useful class of finite, bit-precise C reachability-safety tasks.
The current boundaries further clarify how BTOR2-based program verification can
be extended toward broader C semantics.

%% file: src/6-conclusion.tex
\section{Conclusion}
\label{sec:conclusion}

This paper presented \tool{}, a C-to-BTOR2 verification route for assertion-based reachability safety.
\tool{} models the program into a BTOR2 transition system by encoding control flow with a program counter, representing program data with bit-vectors and arrays, and mapping assumptions and assertion violations to BTOR2 constraints and bad states.
The evaluation shows that this route is effective in practice: \tool{} generates BTOR2 models for 96.7\% of the standard ReachSafety tasks and obtains correct verdicts for 263 cases, while keeping model generation distinct from verification correctness.
It is particularly effective on BV-heavy programs, where it solves 75.5\% of the tasks without wrong verdicts, and it also provides complementary solving capability over CBMC-BMC and Eldarica on functional correctness and restricted pointer-consistency assertions.
The backend experiments further show that C2Btor-generated models can directly benefit from hardware model-checking algorithms, especially k-induction, IC3/PDR, and portfolio-based proving.
Future work will extend this route with more compact memory encodings, richer alias and heap modeling, stronger library summaries, broader C semantic coverage, and more adaptive backend selection.